\documentclass{Interspeech2024}

\interspeechcameraready 
\usepackage{multirow}
\usepackage{makecell}
\usepackage[]{authblk}

\title{Developing vocal system impaired patient-aimed voice quality assessment approach using ASR representation-included multiple features \vspace{-1em}}
\name[affiliation={1}]{Shaoxiang}{Dang}
\name[affiliation={1}]{Tetsuya}{Matsumoto}
\name[affiliation={2}]{Yoshinori}{Takeuchi}
\name[affiliation={3}]{Takashi}{Tsuboi}
\name[affiliation={4}]{Yasuhiro}{Tanaka}
\name[affiliation={5}]{Daisuke}{Nakatsubo}
\name[affiliation={5}]{Satoshi}{Maesawa}
\name[affiliation={5}]{Ryuta}{Saito}
\name[affiliation={3,6}]{Masahisa}{Katsuno}
\name[affiliation={1}]{Hiroaki}{Kudo}

\address{
  $^1$Graduate School of Informatics, Nagoya University\\ $^2$School of Informatics, Daido University\\
  $^3$Department of Neurology, Nagoya University Graduate School of Medicine \\
  $^4$Department of Health Science, Faculty of Psychological and Physical Science, Aichi Gakuin University \\
  $^5$Department of Neurosurgery, Nagoya University Graduate School of Medicine \\
  $^6$Department of Clinical Research Education, Nagoya University
Graduate School of Medicine
  }
\email{dang.shaoxiang.s0@s.mail.nagoya-u.ac.jp}


\begin{document}
\maketitle
\begin{abstract}
    
The potential of deep learning in clinical speech processing is immense, yet the hurdles of limited and imbalanced clinical data samples loom large. This article addresses these challenges by showcasing the utilization of automatic speech recognition and self-supervised learning representations, pre-trained on extensive datasets of normal speech. This innovative approach aims to estimate voice quality of patients with impaired vocal systems. Experiments involve checks on PVQD dataset, covering various causes of vocal system damage in English, and a Japanese dataset focusing on patients with Parkinson's disease before and after undergoing subthalamic nucleus deep brain stimulation (STN-DBS) surgery. The results on PVQD reveal a notable correlation ($>$0.8 on PCC) and an extraordinary accuracy ($<$0.5 on MSE) in predicting Grade, Breathy, and Asthenic indicators. Meanwhile, progress has been achieved in predicting the voice quality of patients in the context of STN-DBS.
    
\end{abstract}

\section{Introduction}

Auditory-perceptual judgment is the primarily subjective approach for assessing the vocal system condition in clinical settings \cite{1,2}. It requires experienced speech pathologists or doctors to comprehensively evaluate sustained vowels and running speech adhering to GRBAS scale. GRBAS scale is a hoarseness evaluation method, it specifically refers to Grade (equivalent to overall severity), Rough, Breathy, Asthenic, and Strained five aspects \cite{3}. Auditory-perceptual judgment is a key means of revealing signs of vocal pathology and monitoring speech disorders after intrusive treatments, for example, subthalamic nucleus deep brain stimulation (STN-DBS) for Parkinson's disease (PD) \cite{4,5}. Auditory-perceptual judgment, however, presents three inconveniences: first, raters are required to possess extensive clinical expertise; second, to enhance the rating reliability, multiple raters are necessarily involved in judgment; third, the lengthy evaluation cycle prevents physicians from promptly obtaining results. These considerations highlight the necessity of objective estimation.

Adhering to the principles valued by speech pathologists, the majority of prior work of objective estimation has focused on predicting using sustained vowels \cite{6,7}. However, due to the limited richness of vowels, it is insufficient for deep learning models that often rely on a large amount of data to acquire robust features. To this end, evaluations of running speech are being considered \cite{8,9,10}, and \cite{10} strongly demonstrates that deep neural gains achieve higher accuracy when trained on continuous speech.

An additional longstanding challenge in shifting auditory-perceptual judgment towards objective assessment analysis lies in the scarcity and imbalance of clinical data, making it difficult for deep learning to extract useful features. Self-supervised learning (SSL) is an efficient unsupervised pre-training technique that has achieved success in many fields \cite{11,12,13,14,15,16}. The research \cite{10} that investigates pre-training via SSL, and subsequently using it as a feature extractor, has effectively provided valuable insights for addressing this issue.


During data analysis, we noticed that comprehension is often relevant to the quality of patient speech. In particular, we hypothesize a positive correlation between automatic speech recognition (ASR) accuracy and speech quality \cite{17,18}. Taking into account the non-linear perception of frequency in auditory-perceptual judgment, we propose a model utilizing mel-spectrogram, ASR representation, and self-supervised learning (SSL) representation for clinical speech quality assessment \cite{19,20}. Unlike previous studies concentrating on a single indicator in GRBAS scale \cite{10,21,22}, this article also delves into an estimation of all GRBAS indicators. Additionally, we conduct experiments to explore the Grade in patients of native Japanese undergoing STN-DBS. Based on experimental results, the proposed method demonstrates improved accuracy and robust predictions.


\section{Proposed methods}
\begin{figure}[t]
\setlength{\abovecaptionskip}{0cm}
  \centering
  \includegraphics[width=\linewidth]{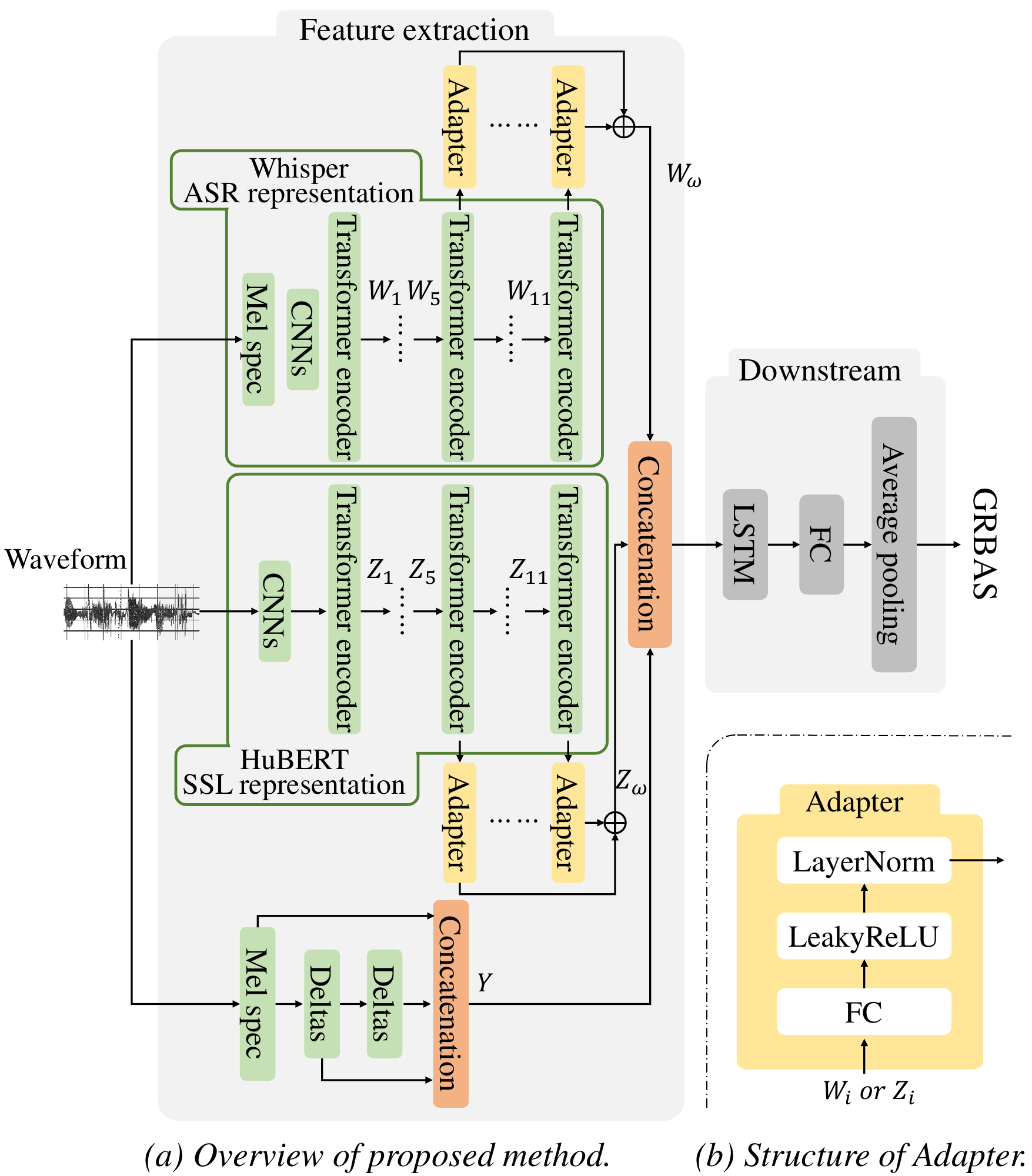}
  \caption{Schematic diagram of the proposed method.}
  \label{fig:1}
\end{figure}

The proposed model consists of two parts: feature extraction and downstream modules. Feature extraction is responsible for obtaining features of ASR representation, SSL representation, and mel-spectrogram, the latter module maps features to GRBAS scales. Fig. \ref{fig:1} illustrates an overview of the proposed method. 


\subsection{Feature extraction}
We employ three types of features for voice quality prediction. The first two are ASR and SSL representations, each pre-trained by the Whisper and HuBERT models respectively \cite{19,23}. Both models stack 12 transformer encoder layers. To fully utilize representations at different depths, representations from the deeper 6 layers of each model undergoes an adapter architecture.

The adapter architecture comprises a fully connected layer, a LeakyReLU layer, and a LayerNorm layer. The outputs of the six adapters are assigned learnable weights that collectively sum to 1, and then these weighted outputs are added together. The third feature is the mel-spectrogram concatenated with its first and second-order deltas. The mel-spectrogram feature is used to represent frequency-level features in auditory-perceptual judgment. HuBERT is pre-trained through SSL with quantized MFCC features as the target, better aligning with auditory characteristics. Unlike HuBERT which directly models time-domain data, Whisper is based on 80-dimensional mel-spectrograms. All three features are input to the downstream module by concatenating them along the feature dimension.

\subsection{Downstream}
It is a consensus that the downstream module does not necessarily need to be very complex when using pre-trained models for feature extraction. Therefore, in the downstream model we designed, we employ two long short-term memory (LSTM) layers for sequence processing, followed by a fully connected layer to map the feature dimensions to the GRBAS dimensions. Finally, we perform average pooling along the time dimension.

\section{Experiment}

\subsection{Datasets}
During the validation of the proposed model, two datasets are used: Perceptual Voice Qualities Database (PVQD) \cite{24} and STN-DBS \cite{4}. 

\subsubsection{PVQD}
PVQD owns 296 speech samples consisting of sustained vowels /a/, /i/, and running speech in English, with each sample representing an individual case (A small number of samples do not contain the sustained vowel /a/ or /i/). Following the methodology of previous studies, sustained vowel /a/ and running speech are extracted separately to create the PVQD-A and PVQD-S datasets \cite{10}. Regarding running speeches, we randomly segment them into speech segments that last 2 to 4 seconds. Employing the same criteria, we divide PVQD-S and PVQD-A into test, validation, and training sets. For instance, the running speech and sustained vowel data of certain patient X are included in the training set of PVQD-S and PVQD-A, respectively. Owing to the high correlation among raters, the averaged opinions of all raters on  the GRBAS are used as labels. Consequently, PVQD data is trained in a regression fashion. In an effort to promote the growth of the clinical speech quality assessment community, we have released the production scripts for PVQD-S and PVQD-A on GitHub\footnote{\url{https://github.com/MydasTouch/PVQD}}.

\subsubsection{STN-DBS}
STN-DBS is a common treatment for PD. Vocal disorder is a recognized side effect of STN-DBS. STN-DBS dataset is collected to assess the degree of vocal disorder \cite{4}. The STN-DBS dataset comprises a total of 96 cases of native Japanese individuals. For simplicity, the recording time (e.g., before or three months after surgery) is not considered. Only vowels are used as input, and the Grade indicator serves as the output. Furthermore, the averaged opinion of raters on Grade is categorized into three intervals: mild for [0,1], moderate for (1,2], and severe for (2,3]. Each case's vocalization includes multiple occurrences of /a/ followed by each of /i/, /u/, /e/, and /o/. In the pre-processing phase, a vocalization utterance is formed by combining one /a/ with /i/, /u/, /e/, and /o/. Utterances from the same patient share a single score. This pre-processing method serves two purposes: the combination of a single /a/ with other vowels, as the vowel /a/ better highlights the patient's vocal cord state. From deep learning's perspective, it helps in expanding the dataset.



\begin{table}[t]
\setlength{\abovecaptionskip}{0cm}
  \caption{Datasets configuration.}
  \label{table:1}
  \centering
  \begin{tabular}{ l rr rr rr  }
    \toprule
    \multirow{2}{*}{\textbf{Datasets}} & \multicolumn{2}{c}{\textbf{Training}} & \multicolumn{2}{c}{\textbf{Validation}} & \multicolumn{2}{c}{\textbf{Testing}} \\
    \cmidrule(l){2-3} \cmidrule(l){4-5} \cmidrule(l){6-7} 
    & \textbf{Utt.}  & \textbf{Pat.}  & \textbf{Utt.}  & \textbf{Pat.} & \textbf{Utt.}  & \textbf{Pat.} \\
    \midrule
     PVQD-S   &    945 & 180  &  305 & 54 & 313   & 60    \\
     PVQD-A   &    196 & 180  &   58 & 53 &  66   & 56            \\
     STN-DBS      &    247 & 60   &   79 & 20 &  71   & 17            \\
    \bottomrule
  \end{tabular}
  
\end{table}

\begin{table*}[th]
\setlength{\abovecaptionskip}{0cm}
\caption{Grade prediction results of proposal and previous models on PVQD-S and PVQD-A datasets.}
  \label{table:2}
    \centering
\resizebox{\textwidth}{!}{
\begin{tabular}{l ccc ccc ccc ccc}
\toprule
\multirow{3}{*}{\textbf{Models}}& \multicolumn{6}{c}{\textbf{PVQD-S}}& \multicolumn{6}{c}{\textbf{ PVQD-A}}\\
\cmidrule(l){2-7} \cmidrule(l){8-13}
& \multicolumn{3}{c}{ \textbf{Utterance Level}}& \multicolumn{3}{c}{\textbf{Patient Level}}& \multicolumn{3}{c}{\textbf{Utterance Level}}& \multicolumn{3}{c}{\textbf{Patient Level}}\\
\cmidrule(l){2-4} \cmidrule(l){5-7} \cmidrule(l){8-10} \cmidrule(l){11-13}
&\textbf{MSE} $\downarrow$& \textbf{PCC} $\uparrow$ & \textbf{SRCC} $\uparrow$  &\textbf{MSE} $\downarrow$& \textbf{PCC} $\uparrow$& \textbf{SRCC} $\uparrow$& \textbf{MSE} $\downarrow$& \textbf{PCC} $\uparrow$& \textbf{SRCC} $\uparrow$&\textbf{MSE} $\downarrow$& \textbf{PCC} $\uparrow$& \textbf{SRCC} $\uparrow$\\
\midrule   

wav2vec2 \cite{10}  & 0.239  & 0.882 & 0.762  & 0.188  & 0.900 & 0.791 & 0.422  & 0.663 & 0.620 & 0.429  & 0.687 &0.694 \\


wav2vec2 + FFNet \cite{27} & 0.208  & 0.887 & 0.782  & 0.166  & 0.901 & 0.802 
& 0.427 & 0.753 & \textbf{0.706}  & 0.445  & 0.766 &\textbf{0.749}\\

OAPVNet \cite{10} & 0.201   & 0.887 & 0.794 & 0.141 &\textbf{0.912} & 0.817 &0.361& 0.742 &0.645 & 0.393  &0.745  &0.675 \\
\midrule 
Proposal & $\textbf{0.171}_{\pm0.302}$   & \textbf{0.892} & \textbf{0.806} & $\textbf{0.136}_{\pm0.228}$ & 0.908 & \textbf{0.827} &$\textbf{0.264}_{\pm0.436}$& \textbf{0.777} &0.679 & $\textbf{0.276}_{\pm0.459}$  & \textbf{0.790}  &0.746 \\
\bottomrule
\end{tabular}}
\end{table*}

\begin{table*}[th]
\setlength{\abovecaptionskip}{0cm}
\caption{GRBAS prediction results of proposal on PVQD-S and PVQD-A datasets.}
  \label{table:3}
    \centering
\resizebox{\textwidth}{!}{
\begin{tabular}{l c ccc ccc ccc ccc}
\toprule
\multirow{3}{*}{\textbf{GRBAS}}& \multirow{3}{*}{ \makecell[c]{\textbf{Intraclass} \\ \textbf{Correlations }  \\  \cite{15}}}&  \multicolumn{6}{c}{\textbf{PVQD-S}}& \multicolumn{6}{c}{\textbf{ PVQD-A}}\\
\cmidrule(l){3-8} \cmidrule(l){9-14}
&& \multicolumn{3}{c}{ \textbf{Utterance Level}}& \multicolumn{3}{c}{\textbf{Patient Level}}& \multicolumn{3}{c}{\textbf{Utterance Level}}& \multicolumn{3}{c}{\textbf{Patient Level}}\\
\cmidrule(l){3-5} \cmidrule(l){6-8} \cmidrule(l){9-11} \cmidrule(l){12-14}
&& \textbf{MSE} $\downarrow$& \textbf{PCC} $\uparrow$ & \textbf{SRCC} $\uparrow$  &\textbf{MSE} $\downarrow$& \textbf{PCC} $\uparrow$& \textbf{SRCC} $\uparrow$& \textbf{MSE} $\downarrow$& \textbf{PCC} $\uparrow$& \textbf{SRCC} $\uparrow$&\textbf{MSE} $\downarrow$& \textbf{PCC} $\uparrow$& \textbf{SRCC} $\uparrow$\\
\midrule   

Grade (G) &0.905  & $0.192_{\pm0.292}$  & 0.881 & 0.773  & $0.160_{\pm0.223}$  & 0.895 & 0.798 
& $0.313_{\pm0.474}$  & 0.730 & 0.583 & $0.325_{\pm0.493}$  & 0.750 &0.662 \\
Rough (R) &0.846 & $0.266_{\pm0.391}$   & 0.707 & 0.604 & $0.237_{\pm0.359}$  & 0.717 & 0.635 
& $0.263_{\pm0.444}$ & 0.614 & 0.514  & $0.287_{\pm0.474}$  & 0.626 & 0.583\\

Breathy (B) & 0.884& $0.189_{\pm0.378}$   & 0.864 & 0.741 &  $0.158_{\pm0.297}$  & 0.880 & 0.750 
& $0.302_{\pm0.583}$ & 0.726 & 0.543  & $0.338_{\pm0.623}$  & 0.737 &0.622\\

Asthenic (A) & 0.892& $0.124_{\pm0.255}$  & 0.894 & 0.803  &$0.104_{\pm0.202}$  & 0.900 & 0.832 
& $0.249_{\pm0.510}$ & 0.745 & 0.492  & $0.281_{\pm0.545}$ & 0.749 &0.540\\

Strained (S)  & 0.862 & $0.455_{\pm0.840}$&  0.348 & 0.404 &$0.423_{\pm0.798}$  & 0.411 & 0.461 
& $0.377_{\pm0.862}$& 0.360 & 0.376  & $0.439_{\pm0.922}$  & 0.323 &0.339\\
\bottomrule
\end{tabular}}
\end{table*}

\subsection{Model parameters}

We employ Whisper\footnote{\url{https://huggingface.co/openai/whisper-base}}, pre-trained on ASR task, to extract ASR features \cite{19}. It's worth noting that Whisper has a padding operation, which is deemed ineffective in current scope. As a result, this padding is removed before utilizing ASR features. For SSL features, we choose the HuBERT\footnote{\url{https://huggingface.co/facebook/hubert-base-ls960}} pre-trained on LibriSpeech \cite{25} for PVQD experiments, and pre-trained HuBERT\footnote{\url{https://huggingface.co/rinna/japanese-hubert-base}} for STN-DBS. Both HuBERT and Whisper consist of 12 encoder layers. The FC layer in adapter transforms features from 768-dimension to 120-dimension. The coefficient of LeakyReLU is set to 0.05. The outputs of adapters, before being applied, undergo a softmax layer to ensure their sum equals 1. The downstream model's FC layer owns 1 or 5 output neurons (depending on single-task or multi-task learning), with each neuron corresponding to one of the five GRABS scales. Also, the number of output neurons of FC layer in downstream is 3, corresponding to the three categories of Grade. The source code for implementing proposed method and some example predictions are available at GitHub\footnote{\url{https://github.com/MydasTouch/GRBASAssessment}}.


\subsection{Training parameters}
\subsubsection{Loss functions}
Due to the high consensus among raters in PVQD, we use mean absolute error (MAE) loss for regression learning. For STN-DBS, drawing inspiration from \cite{26}, we propose a simplified class distance-weighted cross-entropy (SCDW-CE) loss:

\begin{align}
 {\rm SCDW\text{-}CE} = - \log(\hat{y}_c) \times |i-c| 
\end{align}
where $i$ and $c$ denote class index of estimation and ground-truth, respectively. $\hat{y}_c$ represents the probability of predicting the correct label. $|\cdot|$ denotes taking the absolute value.
\subsubsection{Training}
We hire stochastic gradient descent (SGD) as optimizer. The initial learning rate is set to 0.0001, and if there is no improvement on the validation set for four consecutive epochs, the learning rate will be halved. The batch size is 1. During fine-tuning, the weights of the pre-trained modules are not frozen.

\subsubsection{Evaluation metrics}
We use mean squared error (MSE), Pearson correlation coefficient (PCC), and Spearman rank correlation coefficient (SRCC) metrics to assess the performance of regression models. Additionally, for classification modeling, we report Recall, Precision, and F1 scores. Both PVQD and STN-DBS undergo pre-processing where data from a patient is segmented into multiple utterances. Hence, we conduct checks at both the utterance and patient levels. During calculation of patient level, in PVQD, the prediction results for all utterances belonging to the same patient are averaged, while in STN-DBS, the mode is taken for all utterances of a patient, and in cases of multiple modes, the one closest to the mean is selected.

\section{Results and discussions}

\subsection{Results on PVQD}

\begin{figure}[t]
\setlength{\abovecaptionskip}{0cm}
  \centering
  \includegraphics[width=\linewidth]{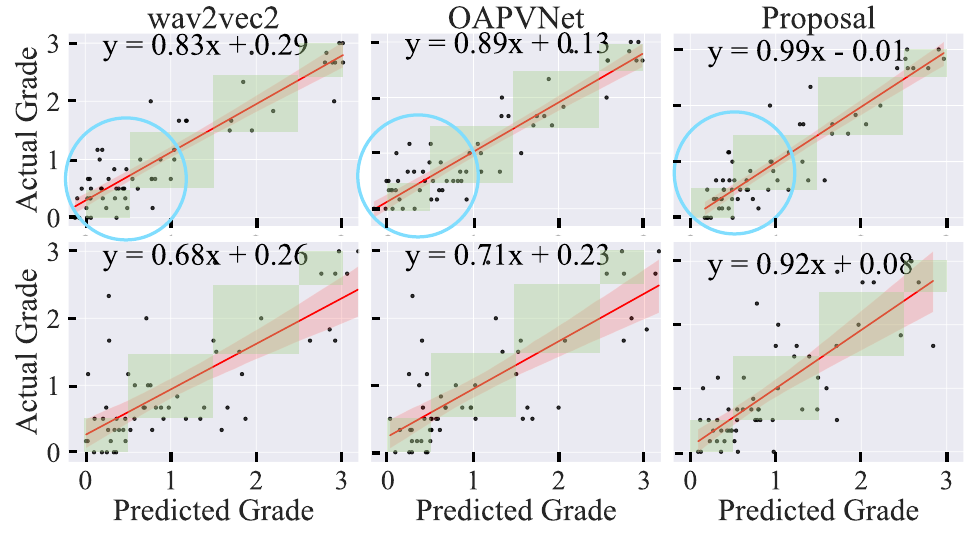}
  \caption{Scatter plots of Grade prediction of patient level on PVQD-S (top row) and PVQD-A (bottom row). The red lines and red shaded areas represent the regression lines and their 95\% confidence interval. The green shadows display the region of error less than 0.5 during auditory-perceptual judgment when discrete scores are rated.}
  \label{fig:2}
\end{figure}

\subsubsection{Results of Grade prediction}
Consistent with prior research, the results of solely predicting Grade are shown in Table \ref{table:2}. In Grade prediction from running speech, all metrics show improved results except for a slight decrease in PCC at the patient level. The metric MSEs, in terms of both utterance and patient levels, show a reduction of 14.9\% and 3.5\%, respectively, and SRCC metrics increase by 1.5\% and 1.2\%, respectively. The results of the experiment predicting Grade using vowels are more significant. MSE decreases by 26.87\% and 29.77\% at the segment and patient levels, respectively.

The prediction results are intuitively presented in Fig. \ref{fig:2}. Each column represents the results of a specific model. When using speech, the proposal's regression equation exhibits a close approximation to the ground truths, with a slope of 0.99 and an intercept of -0.01. The proposed model excels in predicting the audio quality of mild speech (with smaller Grade scores) outperforming previous models, as depicted by the aquamarine circles. This improvement is likely attributed to the newly introduced ASR and Mel features, which benefit the model's ability for more accurate predictions, especially in normal speeches.

In our pursuit of understanding feature effectiveness, we conduct feature ablation studies, and the results are outlined in Table \ref{table:4}. The model attains optimal results by simultaneously utilizing all three features, albeit with a slight decrease in PCC. Additionally, ASR features prove to be more effective for sound quality assessment compared to SSL features. We argue that the superiority is granted by the semantic information involved in ASR pre-training.


\begin{table}[t]
\setlength{\abovecaptionskip}{0cm}
\caption{Results of feature ablation studies.}
  \label{table:4}
    \centering
\resizebox{0.45\textwidth}{!}{
\begin{tabular}{ c cccccccc }
\toprule

\multirow{3}{*}{\textbf{Features}}& \multicolumn{4}{c}{\textbf{PVQD-S}} & \multicolumn{4}{c}{\textbf{PVQD-A}} \\
 \cmidrule(l){2-5} \cmidrule(l){6-9}
& \multicolumn{2}{c}{\textbf{Utt. level}} &  \multicolumn{2}{c}{\textbf{Pat. level}}& \multicolumn{2}{c}{\textbf{Utt. level}} &  \multicolumn{2}{c}{\textbf{Pat. level}} \\
 \cmidrule(l){2-3} \cmidrule(l){4-5} \cmidrule(l){6-7} \cmidrule(l){8-9}
& \textbf{MSE} & \textbf{PCC} & \textbf{MSE} & \textbf{PCC} & \textbf{MSE} & \textbf{PCC} & \textbf{MSE} & \textbf{PCC} \\
\midrule 
ASR + SSL + Mel & \textbf{0.171} & 0.892 & \textbf{0.136} & 0.908 & \textbf{0.264} & \textbf{0.777} & \textbf{0.276} & \textbf{0.790}\\
\midrule 
ASR + Mel & 0.182 & 0.889 & 0.150 & 0.904 & 0.297 & 0.746  & 0.319 &0.757\\
ASR & 0.188 & 0.886 & 0.152 &0.901 & 0.472 & 0.632 & 0.502 &0.673 \\
\midrule 
SSL + Mel & 0.183 &  \textbf{0.894} & 0.146 &  \textbf{0.910} & 0.562 & 0.561 & 0.632 &0.576 \\
SSL & 0.202 & 0.876 & 0.165 &0.892& 0.589 & 0.552 & 0.667 &0.555 \\
\bottomrule
\end{tabular}}
\end{table}

\subsubsection{Results of GRBAS prediction}
The results for predicting all five GRBAS indicators are detailed in Table \ref{table:3}. Firstly, in multi-task learning, the results for predicting Grade using running speech persistently outperform those of previous studies, with the error consistently controlled within a range of 0.5 on MSE. Concurrently, the results of using running speech for BA indicators demonstrate robust positive PCC values (all exceeding 0.8) and SRCC values (all exceeding 0.7). Despite the less abundant information compared to running speech, predictions on vowels yield superior results than the baselines, with almost all correlations surpassing 0.5 and MSEs hovering around 0.3. However, the results for R and S are less promising, indicating lower PCC and higher MSEs. Meanwhile, the second column presents inter-class correlations among raters during scoring, with R and S being the two lowest items. We speculate that due to the lower reliability of R and S during scoring, the labels may less accurately reflect the true condition of patients.

\begin{figure}[t]
\setlength{\abovecaptionskip}{0cm}
  \centering
  \includegraphics[width=\linewidth]{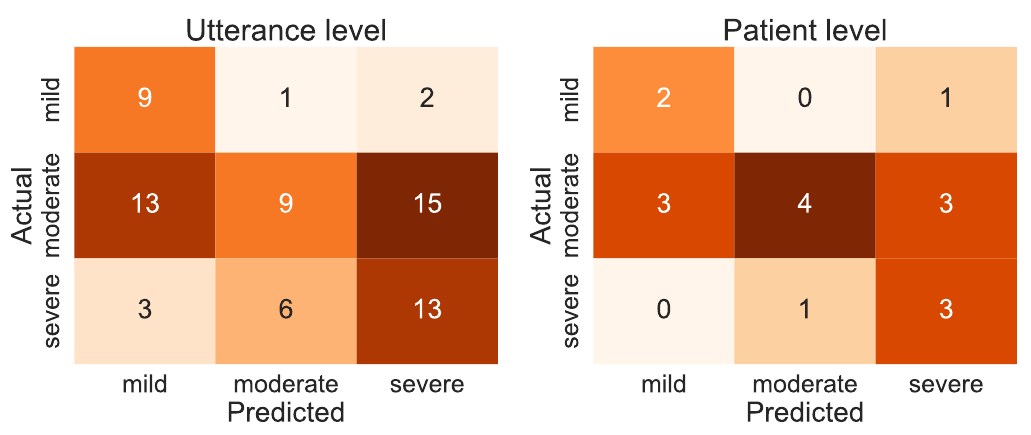}
  \caption{Confusion matrix of predicting Grade on STN-DBS.}
  \label{fig:3}
\end{figure}

\subsection{Results on STN-DBS}

The Grade prediction results on the STN-DBS dataset using all vowels are presented in Table \ref{table:5}. The accuracy of utterance level and patient level are 0.437 and 0.529 respectively. From the confusion matrix in Fig. \ref{fig:3}, we can infer that the model can effectively distinguish extreme samples, such as mild and severe cases. However, for relatively similar samples (closer in distance), errors become significant, such as between mild and moderate, as well as moderate and severe cases.

\begin{table}[t]
\setlength{\abovecaptionskip}{0cm}
\caption{Prediction results of proposal on STN-DBS dataset.}
  \label{table:5}
    \centering
\resizebox{0.38\textwidth}{!}{
\begin{tabular}{ c cc ccc }
\toprule
\multirow{2}{*}{\textbf{Metrics}}

 &\multicolumn{2}{c}{\textbf{Utterance Level}} & \multicolumn{2}{c}{\textbf{Patient Level} }  \\
 \cmidrule(l){2-3} \cmidrule(l){4-5}
  & \textbf{Macro} & \textbf{Weighted} & \textbf{Macro} & \textbf{Weighted}  \\
\midrule   

\textbf{Precision} & 0.452  & 0.488  & 0.543 & 0.642\\
\textbf{Recall}&0.528  & 0.437 & 0.606& 0.529 \\
\textbf{F1 score}& 0.442 &0.414  & 0.526 & 0.530 \\
\midrule
\textbf{Accuracy}&\multicolumn{2}{c}{0.437}&\multicolumn{2}{c}{0.529}  \\
\bottomrule
\end{tabular}}
\end{table}

\subsection{Visualization}
In Fig. 4(a), we present a case from PVQD where, subjectively, the speech signal seems non-pathological, yet its label is 0.5. The OAPVNet, lacking ASR and frequency features, predicts an average of -0.02. In contrast, proposal, incorporating ASR and Mel features, scores 0.39, aligning closer with the label indicating mild symptoms. Notably, this distinction is more evident in the /a/ case, as evidenced by a more sensitive predicted score of 0.82. Fig. 4(b) illustrates an instance labeled as severe but predicted as mild from STN-DBS. Spectrogram analysis suggests suboptimal patient conditions, indicating a loss of normal vocal system functionality (under normal conditions, the patient should pronounce continuous and uninterrupted vowels). This impedes the effective capture of meaningful information. In summary, the physical and mental states of patients pose a formidable challenge in clinical speech quality assessment.
\begin{figure}[t]
\setlength{\abovecaptionskip}{0cm}
  \centering
  \includegraphics[width=\linewidth]{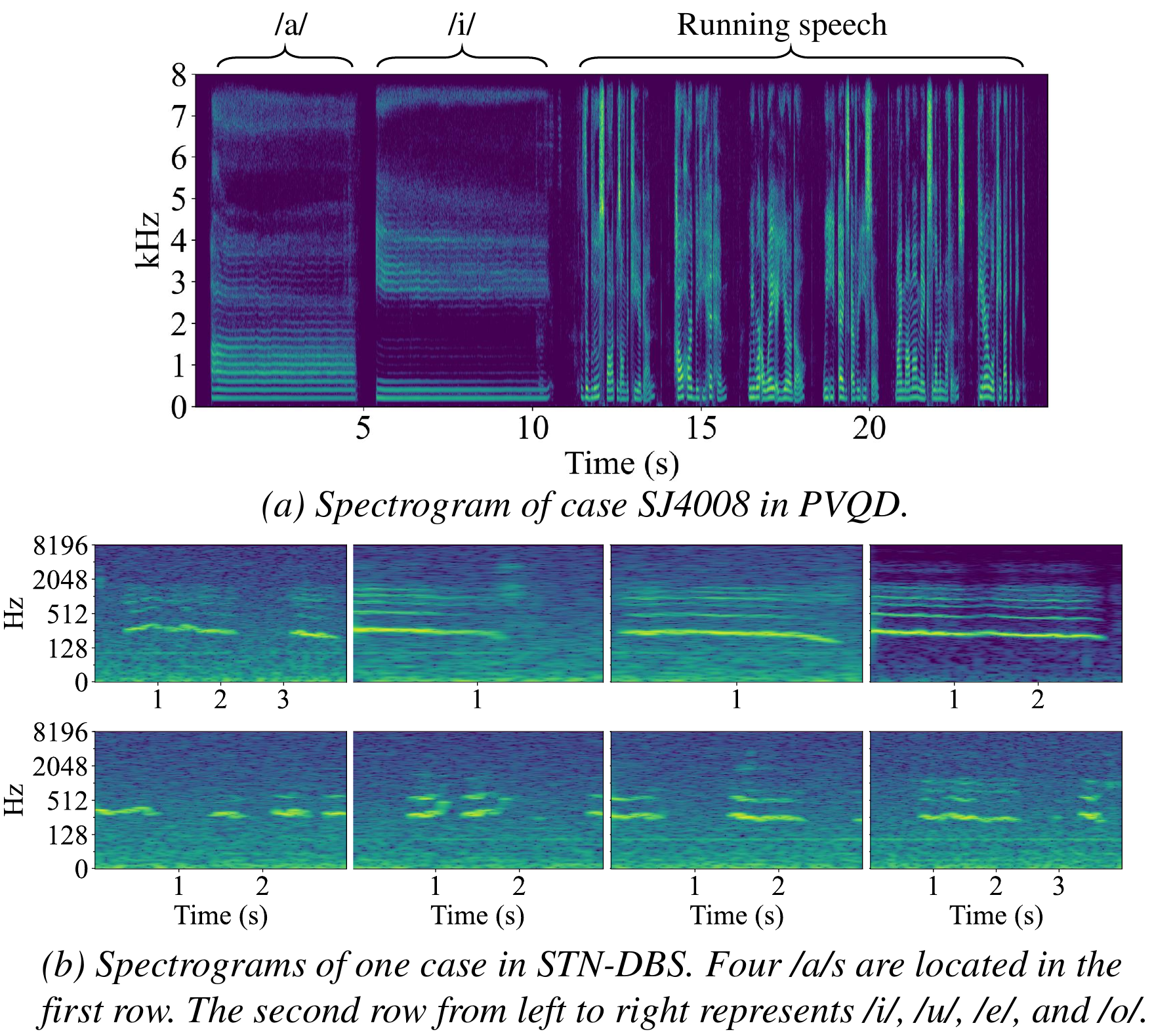}
  \caption{Visualization from PVQD (a) and STN-DBS (b).}
  \label{fig:4}
\end{figure}

\subsection{Discussions}
This paper introduces ASR pre-trained features into clinical speech quality assessment, demonstrating superior performance compared to conventional SSL features. Most importantly, joint usage of ASR, SSL, and Mel features exhibits finer control, especially for normal and mild voices. The benefit of utilizing information-rich running speech is reaffirmed once again. However, voices, in essence, have been used to judge vocal system states so far are indirect information. Therefore, a promising future approach is multi-modal learning that incorporates direct information, such as joint use of perturbation data \cite{28} or laryngoscopic images \cite{29}.


\section{Conclusion}

This article introduces a novel method that integrates ASR, SSL, and mel-spectrogram features for clinical voice quality assessment. The model demonstrates improved accuracy, yielding smaller errors on PVQD-S and PVQD-A datasets. Departing from prior studies that solely predict the super-class Grade, our investigation extends to assess across all GRBAS indicators. The findings strongly indicate that the proposal can achieve comparable and even superior accuracy compared to subjective ratings. Notably, we have practically applied the model to patients with PD who are about to undergo or have undergone STN-DBS, showcasing the preliminary predictive capabilities of the proposed method.

\section{Acknowledgements}
This work is supported by a fellowship of the Nagoya University CIBoG WISE program from MEXT.

\bibliographystyle{IEEEtran}
\bibliography{mybib}

\begin{thebibliography}{10}
\providecommand{\url}[1]{#1}
\csname url@samestyle\endcsname
\providecommand{\newblock}{\relax}
\providecommand{\bibinfo}[2]{#2}
\providecommand{\BIBentrySTDinterwordspacing}{\spaceskip=0pt\relax}
\providecommand{\BIBentryALTinterwordstretchfactor}{4}
\providecommand{\BIBentryALTinterwordspacing}{\spaceskip=\fontdimen2\font plus
\BIBentryALTinterwordstretchfactor\fontdimen3\font minus \fontdimen4\font\relax}
\providecommand{\BIBforeignlanguage}[2]{{%
\expandafter\ifx\csname l@#1\endcsname\relax
\typeout{** WARNING: IEEEtran.bst: No hyphenation pattern has been}%
\typeout{** loaded for the language `#1'. Using the pattern for}%
\typeout{** the default language instead.}%
\else
\language=\csname l@#1\endcsname
\fi
#2}}
\providecommand{\BIBdecl}{\relax}
\BIBdecl

\bibitem{1}
B.~Barsties and M.~De~Bodt, ``Assessment of voice quality: Current state-of-the-art,'' \emph{Auris Nasus Larynx}, vol.~42, no.~3, pp. 183--188, 2015.

\bibitem{2}
J.~Kreiman and B.~R. Gerratt, ``Perceptual assessment of voice quality: Past, present, and future,'' \emph{Perspectives on Voice and Voice Disorders}, vol.~20, no.~2, pp. 62--67, 2010.

\bibitem{3}
M.~Hirano, ``Psyco-acoustic evaluation of voice,'' \emph{Clinical examination of voice: disorders of human communication}, pp. 81--84, 1981.

\bibitem{4}
T.~Tsuboi, H.~Watanabe, Y.~Tanaka, R.~Ohdake, N.~Yoneyama, K.~Hara, R.~Nakamura, H.~Watanabe, J.~Senda, N.~Atsuta \emph{et~al.}, ``Distinct phenotypes of speech and voice disorders in parkinson's disease after subthalamic nucleus deep brain stimulation,'' \emph{Journal of Neurology, Neurosurgery \& Psychiatry}, vol.~86, no.~8, pp. 856--864, 2015.

\bibitem{5}
T.~Tsuboi, H.~Watanabe, Y.~Tanaka, R.~Ohdake, M.~Hattori, K.~Kawabata, K.~Hara, M.~Ito, Y.~Fujimoto, D.~Nakatsubo \emph{et~al.}, ``Early detection of speech and voice disorders in parkinson’s disease patients treated with subthalamic nucleus deep brain stimulation: a 1-year follow-up study,'' \emph{Journal of Neural Transmission}, vol. 124, pp. 1547--1556, 2017.

\bibitem{6}
J.~A. G{\'o}mez-Garc{\'\i}a, L.~Moro-Vel{\'a}zquez, J.~Mendes-Laureano, G.~Castellanos-Dominguez, and J.~I. Godino-Llorente, ``Emulating the perceptual capabilities of a human evaluator to map the grb scale for the assessment of voice disorders,'' \emph{Engineering Applications of Artificial Intelligence}, vol.~82, pp. 236--251, 2019.

\bibitem{7}
T.~Schraut, A.~Sch{\"u}tzenberger, T.~Arias-Vergara, M.~Kunduk, M.~Echternach, and M.~D{\"o}llinger, ``Machine learning based estimation of hoarseness severity using sustained vowels,'' \emph{The Journal of the Acoustical Society of America}, vol. 155, no.~1, pp. 381--395, 2024.

\bibitem{8}
C.~Moers, B.~M{\"o}bius, F.~Rosanowski, E.~N{\"o}th, U.~Eysholdt, and T.~Haderlein, ``Vowel-and text-based cepstral analysis of chronic hoarseness,'' \emph{Journal of Voice}, vol.~26, no.~4, pp. 416--424, 2012.

\bibitem{9}
S.~Xie, N.~Yan, P.~Yu, M.~L. Ng, L.~Wang, Z.~Ji \emph{et~al.}, ``Deep neural networks for voice quality assessment based on the grbas scale.'' in \emph{Interspeech}, 2016, pp. 2656--2660.

\bibitem{10}
S.~Dang, T.~Matsumoto, Y.~Takeuchi, H.~Kudo, T.~Tsuboi, Y.~Tanaka, and M.~Katsuno, ``Using self-learning representations for objective assessment of patient voice in dysphonia,'' in \emph{2022 Asia-Pacific Signal and Information Processing Association Annual Summit and Conference (APSIPA ASC)}.\hskip 1em plus 0.5em minus 0.4em\relax IEEE, 2022, pp. 359--363.

\bibitem{11}
Y.~Gao, H.~Shi, C.~Chu, and T.~Kawahara, ``Enhancing two-stage finetuning for speech emotion recognition using adapters,'' in \emph{ICASSP 2024-2024 IEEE International Conference on Acoustics, Speech and Signal Processing (ICASSP)}.\hskip 1em plus 0.5em minus 0.4em\relax IEEE, 2024, pp. 11\,316--11\,320.

\bibitem{12}
Y.~Gao, C.~Chu, and T.~Kawahara, ``{Two-stage Finetuning of Wav2vec 2.0 for Speech Emotion Recognition with ASR and Gender Pretraining},'' in \emph{Proc. INTERSPEECH 2023}, 2023, pp. 3637--3641.

\bibitem{13}
X.~Shi, X.~Li, and T.~Toda, ``Emotion awareness in multi-utterance turn for improving emotion prediction in multi-speaker conversation,'' in \emph{Proc. Interspeech}, vol. 2023, 2023, pp. 765--769.

\bibitem{14}
J.~Tian, D.~Hu, X.~Shi, J.~He, X.~Li, Y.~Gao, T.~Toda, X.~Xu, and X.~Hu, ``Semi-supervised multimodal emotion recognition with consensus decision-making and label correction,'' in \emph{Proceedings of the 1st International Workshop on Multimodal and Responsible Affective Computing}, 2023, pp. 67--73.

\bibitem{15}
S.~Dang, T.~Matsumoto, Y.~Takeuchi, and H.~Kudo, ``{Using Semi-supervised Learning for Monaural Time-domain Speech Separation with a Self-supervised Learning-based SI-SNR Estimator},'' in \emph{Proc. INTERSPEECH 2023}, 2023, pp. 3759--3763.

\bibitem{16}
H.~Sun, S.~Zhao, X.~Wang, W.~Zeng, Y.~Chen, and Y.~Qin, ``Fine-grained disentangled representation learning for multimodal emotion recognition,'' in \emph{ICASSP 2024-2024 IEEE International Conference on Acoustics, Speech and Signal Processing (ICASSP)}.\hskip 1em plus 0.5em minus 0.4em\relax IEEE, 2024, pp. 11\,051--11\,055.

\bibitem{17}
H.~Shi, L.~Wang, S.~Li, C.~Fan, J.~Dang, and T.~Kawahara, ``Spectrograms fusion-based end-to-end robust automatic speech recognition,'' in \emph{2021 Asia-Pacific Signal and Information Processing Association Annual Summit and Conference (APSIPA ASC)}, 2021, pp. 438--442.

\bibitem{18}
T.~Song, Q.~Xu, M.~Ge, L.~Wang, H.~Shi, Y.~Lv, Y.~Lin, and J.~Dang, ``{Language-specific Characteristic Assistance for Code-switching Speech Recognition},'' in \emph{Proc. Interspeech 2022}, 2022, pp. 3924--3928.

\bibitem{19}
A.~Radford, J.~W. Kim, T.~Xu, G.~Brockman, C.~McLeavey, and I.~Sutskever, ``Robust speech recognition via large-scale weak supervision,'' in \emph{International Conference on Machine Learning}.\hskip 1em plus 0.5em minus 0.4em\relax PMLR, 2023, pp. 28\,492--28\,518.

\bibitem{20}
H.~Shi, M.~Mimura, and T.~Kawahara, ``Waveform-domain speech enhancement using spectrogram encoding for robust speech recognition,'' \emph{IEEE/ACM Transactions on Audio, Speech, and Language Processing}, pp. 1--12, 2024.

\bibitem{21}
F.~Jalalinajafabadi, C.~Gadepalli, M.~Ghasempour, M.~Luj{\'a}n, B.~Cheetham, and J.~Homer, ``Computerised objective measurement of strain in voiced speech,'' in \emph{2015 37th Annual International Conference of the IEEE Engineering in Medicine and Biology Society (EMBC)}.\hskip 1em plus 0.5em minus 0.4em\relax IEEE, 2015, pp. 5589--5592.

\bibitem{22}
F.~Jalalinajafabadi, C.~Gadepalli, M.~Ghasempour, F.~Ascott, M.~Luj{\'a}n, J.~Homer, and B.~Cheetham, ``Objective assessment of asthenia using energy and low-to-high spectral ratio,'' in \emph{2015 12th International Joint Conference on e-Business and Telecommunications (ICETE)}, vol.~5.\hskip 1em plus 0.5em minus 0.4em\relax IEEE, 2015, pp. 76--83.

\bibitem{23}
W.-N. Hsu, B.~Bolte, Y.-H.~H. Tsai, K.~Lakhotia, R.~Salakhutdinov, and A.~Mohamed, ``Hubert: Self-supervised speech representation learning by masked prediction of hidden units,'' \emph{IEEE/ACM Transactions on Audio, Speech, and Language Processing}, vol.~29, pp. 3451--3460, 2021.

\bibitem{24}
P.~R. Walden, ``Perceptual voice qualities database (pvqd): Database characteristics,'' \emph{Journal of Voice}, vol.~36, no.~6, pp. 875--e15, 2022.

\bibitem{27}
A.~Ragano, E.~Benetos, M.~Chinen, H.~B. Martinez, C.~K. Reddy, J.~Skoglund, and A.~Hines, ``A comparison of deep learning mos predictors for speech synthesis quality,'' in \emph{2023 34th Irish Signals and Systems Conference (ISSC)}.\hskip 1em plus 0.5em minus 0.4em\relax IEEE, 2023, pp. 1--6.

\bibitem{25}
V.~Panayotov, G.~Chen, D.~Povey, and S.~Khudanpur, ``Librispeech: an asr corpus based on public domain audio books,'' in \emph{2015 IEEE international conference on acoustics, speech and signal processing (ICASSP)}.\hskip 1em plus 0.5em minus 0.4em\relax IEEE, 2015, pp. 5206--5210.

\bibitem{26}
G.~Polat, I.~Ergenc, H.~T. Kani, Y.~O. Alahdab, O.~Atug, and A.~Temizel, ``Class distance weighted cross-entropy loss for ulcerative colitis severity estimation,'' in \emph{Annual Conference on Medical Image Understanding and Analysis}.\hskip 1em plus 0.5em minus 0.4em\relax Springer, 2022, pp. 157--171.

\bibitem{28}
J.~D. Arias-Londo{\~n}o, J.~A. G{\'o}mez-Garc{\'\i}a, and J.~I. Godino-Llorente, ``Multimodal and multi-output deep learning architectures for the automatic assessment of voice quality using the grb scale,'' \emph{IEEE Journal of Selected Topics in Signal Processing}, vol.~14, no.~2, pp. 413--422, 2019.

\bibitem{29}
T.~Tsuboi, H.~Watanabe, Y.~Tanaka, R.~Ohdake, N.~Yoneyama, K.~Hara, M.~Ito, M.~Hirayama, M.~Yamamoto, Y.~Fujimoto \emph{et~al.}, ``Characteristic laryngoscopic findings in parkinson’s disease patients after subthalamic nucleus deep brain stimulation and its correlation with voice disorder,'' \emph{Journal of Neural Transmission}, vol. 122, pp. 1663--1672, 2015.

\end{thebibliography}

\end{document}